\documentclass[11pt, a4paper,notitlepage]{article}

\usepackage{amssymb}
\usepackage{amsmath}
\usepackage{amsthm}
\usepackage{amsfonts}
\usepackage{mathrsfs}
\usepackage{epsfig}

\newcommand{\EE}{\mathbb{E}}

\newcommand{\eps}{\varepsilon}
\newcommand{\dd}{\mathrm{d}}
\newcommand{\FF}{\mathcal{F}}

\newcommand{\RR}{\mathbb{R}}
\newcommand{\NN}{\mathbb{N}}

\newcommand{\ZZ}{\mathbb{Z}}

\newcommand{\qv}[1]{\left<  #1  \right>}

\newcommand{\half}{\frac{1}{2}}

\newcommand{\rr}{\mathbb{R}}
\newcommand{\nn}{\mathbb{N}}

\newcommand{\bx}{\mathbf{x}}
\newcommand{\bu}{\mathbf{u}}
\newcommand{\bv}{\mathbf{v}}
\newcommand{\bw}{\mathbf{w}}

\newcommand{\bs}{\mathbf{s}}
\newcommand{\bk}{\mathbf{k}}
\newcommand{\bX}{\mathbf{X}}

\newcommand{\bbf}{\mathbf{f}}

\newcommand{\ii}{{\rm i}}

\newtheorem{thm}{Theorem}[section]

\newtheorem{cor}[thm]{Corollary}
\theoremstyle{definition}
\newtheorem{rem}[thm]{Remark}
\newtheorem{ex}[thm]{Example}

\newtheorem{cond}[thm]{Condition}

\newenvironment{prf}{\begin{proof}[\bf Proof]}{\end{proof}}

\def\infint{\int_{-\infty}^\infty}
\def\ex{\mathbb{E}}
\def\var{{\rm Var\,}}

\begin{document}

\title{Nonparametric  methods for  volatility density estimation}

\author{Bert van Es, Peter Spreij \\[1ex]
{\normalsize\sl Korteweg-de Vries Institute for Mathematics}\\
{\normalsize\sl Universiteit van Amsterdam}\\
{\normalsize\sl PO Box 94248}\\
{\normalsize\sl 1090GE Amsterdam}\\
{\normalsize\sl The Netherlands}\\[2ex] \and
Harry van Zanten\\[1ex]
{\normalsize\sl Department of Mathematics}\\
{\normalsize\sl Eindhoven University of Technology}\\
{\normalsize\sl PO Box 513}\\
{\normalsize\sl 5600 MB Eindhoven}\\
{\normalsize\sl The Netherlands}}

\maketitle

\begin{abstract}
Stochastic volatility modelling of financial processes has become increasingly popular.
The proposed models usually contain a stationary volatility process.
We will motivate and review several nonparametric methods for estimation of the density of
the volatility process.  Both models based on  discretely sampled continuous time processes
and discrete time models will  be discussed.

The key insight for the analysis is a transformation of the volatility density estimation problem to a
deconvolution model for which standard methods exist.
Three type of nonparametric density estimators are reviewed: the Fourier-type deconvolution kernel density estimator,
 a wavelet deconvolution density estimator and a penalized projection estimator.
The performance of these estimators will be compared.
\end{abstract}

\section{Introduction}

We discuss a number of nonparametric methods that come into play
when one wants to estimate the density of the volatility process,
given observations of the  price process of some asset.
The models that we treat are mainly formulated in continuous time,
although we pay some separate attention to discrete time models.
The observations of the continuous time models will always be in
discrete time however and may occur at low frequency (fixed lag
between observation instants), or high frequency (vanishing time
lag). In this review, for simplicity  we focus on the
univariate marginal distribution of the volatility process,
although similar results can be obtained for multivariate marginal
distributions.
%, see~Van Es et al.~\cite{bert04} and Van Es and
%Spreij~\cite{VanEsSpreij09}.

Although the underlying models differ in the sense that they are
formulated either in continuous or in discrete time, in all cases
the observations are given by a discrete time process. Moreover,
as we shall see, the observation scheme can always (approximately)
be cast as of  `signal plus noise' type
\[
Y_i = X_i + \eps_i,
\]
where $X_i$ is to be interpreted as the `signal'. If for fixed $i$
the random variables $X_i$ and $\eps_i$ are independent, the
distribution of the $Y_i$ is a convolution of the distributions of
$X_i$ and $\eps_i$. The density of the `signal' $X_i$ is the
object of interest, while the density of the `noise' $\eps_i$ is
supposed to be known to the observer. The statistical problem is
to recover the density of the signal by {\em deconvolution}.
Classically for such models it was often also assumed that the
processes $(X_i)$ and $(\eps_i)$ are i.i.d. Under these conditions
Fan~\cite{104} gave lower bounds for the estimation of the unknown
density $f$ at a fixed point $x_0$  and showed that kernel-type
estimators achieve the optimal rate. An alternative estimation
method was proposed in the paper Pensky and Vidakovic~\cite{103},
using wavelet methods instead of kernel estimators and where
global $L^2$-errors were considered instead of pointwise errors.

However, for the stochastic volatility models that we consider,
the i.i.d.\ assumption on the $X_i$ is violated. Instead,  the
$X_i$ may be modelled as  stationary random variables, that are
allowed to exhibit some form of weak dependence, controlled by
appropriate mixing properties,  strongly mixing or $\beta$-mixing.
These mixing conditions are justified by the fact that they are
satisfied for many popular GARCH-type and stochastic volatility
models (see e.g.\ Carrasco and Chen~\cite{Car02}), as well as for
continuous time models, where $\sigma^2$ is solves  a stochastic
differential equation, see e.g.\ Genon-Catalot et
al.~\cite{Gen00}. The estimators that we discuss are based on
kernel methods, wavelets and penalized contrast estimation, also
referred to as penalized projection estimation. We will review the
performance of these deconvolution estimators under weaker than
i.i.d.\ assumptions and show that this essentially depends on the
smoothness and mixing conditions of the underlying process and the
frequency of the observations. For a survey of other nonparametric
statistical problems for financial data  we refer to Franke et
al.~\cite{FKM09}

%Let us mention that Masry \cite{Mas91} studied the properties of the deconvolution kernel estimators under
%various mixing assumptions. In particular, he showed that under a (rather stringent) uniform mixing
%condition, certain kernel estimators still achieve the
%optimal i.i.d.\ rates of \cite{104} in the case that $f$ has two bounded derivatives.
%Under the less demanding strong mixing condition that we consider he obtained a sub-optimal rate.

%To illustrate the main results we apply them to the model (\ref{eq: sv}). The results we find
%in this case extend and complement results from the papers \cite{bert03} and \cite{bert04}.
%In the first of these papers the same model is studied, but it is assumed that the time $\Delta$ between
%the observations vanishes as the number of observations grows. In the present paper we keep
%$\Delta$ fixed (low frequency data).

The paper is organized as follows. In
Section~\ref{section:ctsmodel} we introduce the continuous time model. In Section~\ref{section:kernel} we consider a kernel type estimator of the invariant volatility density and apply it to a set of real data.
Section~\ref{section:wavelet} is devoted to a wavelet density estimator and in Section~\ref{section:ppe}  a minimum contrast estimator is discussed. Some related results for discrete time models are reviewed in
Section~\ref{section:edtm} and Section~\ref{section:remarks} contains some concluding remarks.

\section{The continuous time model}\label{section:ctsmodel}

Let $S$ denote the log price process of some stock in a financial
 market. It is often assumed that $S$ can be modelled as the
solution of a stochastic differential equation or, more general,
as an It\^o diffusion process. So we assume that we can write
\begin{equation}\label{eq:sde}
\dd S_t = b_t\, \dd t + \sigma_t \,\dd W_t, \ \ \ S_0=0,
\end{equation}
or, in integral form,
\begin{equation}\label{eq:si}
S_t = \int_0^t b_s\, \dd s + \int_0^t \sigma_s \,\dd W_s,
\end{equation}
where $W$ is a standard Brownian motion and the processes $b$ and
$\sigma$ are assumed to satisfy certain regularity conditions (see
Karatzas and Shreve~\cite{KS91}) to have  the integrals in
(\ref{eq:si}) well-defined. In a financial context, the process
$\sigma$ is called the volatility process. One often takes the
process $\sigma$ independent of the Brownian motion $W$.

Adopting this common assumption throughout the paper, unless explicitly stated otherwise, we also assume that $\sigma$ is a
strictly stationary positive process satisfying a mixing
condition, for example an ergodic diffusion on $(0,\infty)$. We
will assume that the one-dimensional marginal distribution of
$\sigma$  has an invariant density with respect to the Lebesgue
measure on $(0,\infty)$. This is typically the case in virtually
all stochastic volatility models that are proposed in the
literature, where the evolution of $\sigma$ is modelled by a
stochastic differential equation, mostly in terms of $\sigma^2$,
or $\log \sigma^2$ (cf.\ e.g.\ Wiggins~\cite{Wiggins87},
Heston~\cite{Heston93}). Often $\sigma^2_t$ is a function of a
process $X_t$ satisfying a stochastic differential equation of the
type
\begin{equation}\label{eq:sigmasde}
\dd X_t = b(X_t)\, \dd t + a(X_t)\, \dd B_t,
\end{equation}
with $B_t$ a Brownian motion. Under regularity conditions,
the invariant density of $X$ is up to a multiplicative constant equal to
\begin{equation}
\label{eq: piet} x \mapsto \frac{1}{a^2(x)}\,{\exp
\left(2\int_{x_0}^x \frac{b(y)}{a^2(y)}\,\dd y\right)},
\end{equation}
where $x_0$ is an arbitrary element of the state space, see
 e.g.\ Gihman and Skorohod~\cite{GS72} or Skorokhod~\cite{Skorohod89}. From
formula~(\ref{eq: piet}) one sees that the invariant distribution
of the volatility process (take $X$ for instance equal to
$\sigma^2$ or $\log \sigma^2$) may take  on many different forms,
as is the case for the various models that have been proposed in
the literature. In absence of parametric assumptions on the
coefficients $a$ and $b$, we will investigate
nonparametric procedures to estimate the corresponding densities,
even refraining from an underlying model like~(\ref{eq:sigmasde}),
partly aimed at recovering possible `stylized facts' exhibited by
the observations.

For instance, one could think of {\em volatility
clustering}. This may be cast by saying that for different time
instants $t_1,t_2$ that are close, the corresponding values of
$\sigma_{t_1},\sigma_{t_2}$ are close again. This can partly be
explained by assumed continuity of the process $\sigma$, but it
might also result from specific areas around the diagonal where the multivariate
density of $(\sigma_{t_1},\sigma_{t_2})$ assumes high values if $t_1$ and $t_2$ are relatively close. It
is therefore conceivable that  the density of
$(\sigma_{t_1},\sigma_{t_2})$ has high concentrations around
points $(\ell,\ell)$ and $(h,h)$, with $\ell<h$, a kind of
bimodality of the joint distribution, with the interpretation that
clustering occurs around a low value $\ell$ or around a high value
$h$. This in turn may be reflected by bimodality of the univariate
marginal distribution of $\sigma_t$.

A situation in which this naturally occurs is the following. Consider a regime switching volatility process. Assume that for
$i=0,1$ we have two stationary processes $X^i$  having stationary
densities $f^i$.   We assume these two processes to be
independent, and also independent of a two-state stationary
homogeneous Markov chain $U$ with states $0,1$. The stationary
distribution of $U$ is given by $\pi_i:=P(U_t=i)$. The process
$\xi$ is defined by
\[
\xi_t=U_tX^1_t+(1-U_t)X^0_t.
\]
Then $\xi$ is stationary too and it has a  stationary  density $f$
given by
\begin{align*}
f(x) & = \pi_1f^1(x) + \pi_0f^0(x).
\end{align*}
Suppose that the volatility process is defined by
$\sigma^2_t=\exp(\xi_t)$ and that the $X^i$ are both
Ornstein-Uhlenbeck processes given by
\[
\dd X^i_t = -b_i(X^i_t-\mu_i)\,\dd t + a_i\,\dd W^i_t,
\]
with $W^1$, $W^2$ independent Brownian motions, $\mu_1\neq \mu_2$
and $b_1,b_2>0$. Suppose that the $X^i$ start in their stationary
$N(\mu_i,\frac{a_i^2}{2b_i})$ distributions. Then the stationary
density  $f$ is a bimodal mixture of normal densities with $\mu_1$ and $\mu_2$ as the locations
of the local maxima.
Nonparametric procedures are able to detect such a property and are consequently by all means sensible
tools to get some first insights into the shape of the invariant
density.

A first object of study is the marginal univariate distribution of
the stationary volatility process $\sigma$. The standing
assumption in all what follows is that this distribution admits a
density   w.r.t.\ Lebesgue measure. We will also consider the
invariant density of the integrated squared volatility process
over an interval of length $\Delta$. By stationarity of $\sigma$
this is the density of $\int_0^\Delta \sigma_t^2\,\dd t$. We will
consider density estimators and assess their quality by giving results on
their mean squared or integrated mean squared error. For kernel
estimators, we rely on Van Es et al.~\cite{bert03}, where this
problem has been studied for the marginal {\em univariate} density
of $\sigma$. In~Van Es and Spreij~\cite{VanEsSpreij09} one can
find results for multivariate density estimators. Results on
wavelet estimators will be taken from Van Zanten and
Zareba~\cite{VanzantenZareba08}. Penalized contrast estimators have
been treated in Comte and Genon-Catalot~\cite{com}.

The observations of log-asset price $S$ process are assumed to
take place at the time instants $0, \Delta, 2\Delta, \ldots,
n\Delta$. In case one deals with low frequency observations,
$\Delta$ is fixed. For high frequency observations, the time gap
satisfies $\Delta = \Delta_n \to 0$ as $n \to \infty$. To obtain
consistency for the estimators that we will study in the latter
case, we will make the additional assumption $n\Delta_n \to
\infty$.

To explain the origin of the estimators that we consider in this
paper, we often work with the simplified model, which is obtained
from~(\ref{eq:sde}) by taking $b_t=0$. We then suppose to have
discrete-time data $S_0, S_\Delta, S_{2\Delta}, \ldots$ from a
continuous-time stochastic volatility model of the form
\begin{equation*}
\dd S_t = \sigma_t\,\dd W_t.
\end{equation*}
Under this additional assumption, we will see that we
(approximately) deal with stationary observations $Y_i$ that can
be represented as $Y_i=X_i+\eps_i$, where for each $i$ the random
variables $X_i$ and $\eps_i$ are independent.

\section{Kernel deconvolution}\label{section:kernel}

In this section we consider kernel deconvolution density estimators. We construct them, give expressions for
bias and variance and give an application to real data.

\subsection{Construction of the estimator}

To motivate the construction of the estimator, we first
consider~(\ref{eq:sde}) without the drift term, so we assume to
have the simplified model
\begin{equation}\label{eq:sdesimple}
  \dd S_t = \sigma_t \,\dd W_t, \ \ \ S_0=0.
\end{equation}
It is  assumed that we observe the process $S$
at the discrete time instants $0$, $\Delta$,
$2\Delta, \ldots, n\Delta$, satisfying $\Delta\to 0$, $n\Delta\to\infty$.
For $i=1, 2, \ldots$ we work, as in
Genon-Catalot et al.~\cite{GCJL98,GCJL99}, with  the normalized
increments
\[
X^{\Delta}_i = \tfrac{1}{\sqrt{\Delta}}(S_{i\Delta} -
S_{(i-1)\Delta}).
\]
For small $\Delta$, we have the  rough approximation
\begin{align}
X^{\Delta}_i & = \tfrac{1}{\sqrt{\Delta}}
\int_{(i-1)\Delta}^{i\Delta}\sigma_t\,\dd W_t \nonumber \\
& \approx
\sigma_{(i-1)\Delta}  \tfrac{1}{\sqrt{\Delta}}(W_{i\Delta} -
W_{(i-1)\Delta}) \label{eq:approx}\\
& = \sigma_{(i-1)\Delta}Z^\Delta_i,\nonumber
\end{align}
where for $i=1, 2, \ldots$ we define
\[
Z^\Delta_i= \frac{1}{\sqrt{\Delta}}(W_{i\Delta} -
W_{(i-1)\Delta}).
\]
By the independence and stationarity of Brownian increments, the
sequence $Z^\Delta_1, Z^\Delta_2, \ldots$ is an i.i.d.\ sequence
of standard normal random variables. Moreover,  the sequence is
independent of the process $\sigma$ by assumption.

Writing $Y_i=\log (X_i^{\Delta})^2$,
$\xi_i=\log\sigma_{(i-1)\Delta}^2$, $\eps_i=\log  (Z_i^{\Delta})^2$
and taking the logarithm of the square of
 $X_i^\Delta$  we get
\begin{align*}
Y_i & \approx \xi_i +\eps_i,
\end{align*}
where the terms in the sum are independent. Assuming that the
approximation is sufficiently accurate we can use this approximate
convolution structure to estimate the unknown density $f$ of $\log
\sigma_{i\Delta}^2$ from the transformed observed $Y_i=\log
(X^\Delta_i)^2$. The characteristic functions involved are denoted
by $\phi_Y$, $\phi_\xi$ and $\phi_k$, where $k$ is the density of
the `noise' $\log (Z_i^\Delta)^2$. One obviously has
$\phi_Y=\phi_\xi\phi_k$ and one easily sees that the density $k$ is given by
\[
k(x)= \frac{1}{\sqrt{2\pi}}\, e^{\tfrac{1}{2}x} e^{-
\tfrac{1}{2}e^{x}},
\]
and its characteristic function by
\[
\phi_k(t)= \frac{1}{\sqrt{\pi}}\, 2^{i t} \Gamma\Big(\frac12 + it\Big).
\]

The idea of getting a deconvolution estimator of $f$ is simple.
Using a kernel function $w$, a bandwidth $h$, and the $Y_i$, the
density $g$  of the $Y_i$ is estimated by
\[
g_{nh}(y)=\frac{1}{nh}\sum_jw\big(\frac{y-Y_j}{h}\big).
\]
Denoting $\phi_{g,nh}$ the characteristic function of  $g_{nh}$,
one estimates $\phi_Y$ by $\phi_{g,nh}$ and $\phi_\xi$ by
$\phi_{g,nh}/\phi_k$. Following a well-known approach in
statistical deconvolution theory (see e.g.\ Section 6.2.4 of Wand
and Jones~\cite{WandJones95}), Fourier inversion then yields the
density estimator of $f$. By elementary calculations one obtains
from this procedure
\begin{equation}\label{eq:fnhc}
f_{nh}(x)=\frac{1}{nh}\sum_{j=1}^n
v_h\left(\frac{x-\log(X^\Delta_j)^2}{h}\right),
\end{equation}
where $v_h$ is  the kernel function, depending on the bandwidth
$h$,
\begin{equation}\label{fourkernel}
v_h(x)=\frac{1}{2\pi}\infint \frac{\phi_w(s)}{\phi_k(s/h)}\
e^{-\ii sx}\,\dd s.
\end{equation}
One   easily verifies that the  estimator $f_{nh}$, is real-valued.

To justify the approximation in~(\ref{eq:approx}), we quantify a
stochastic continuity property of $\sigma^2$. In addition to this we make
the mixing condition explicit. We impose
\begin{cond}\label{cond:sigma} The process $\sigma^2$ satisfies the following conditions.
\begin{enumerate}
\item
It is $L^1$-H\"older continuous of order one half,
$\ex|\sigma^2_t -
\sigma_0^2| = O(t^{1/2})$ for $t \to 0$.
\item
It is
strongly mixing with coefficient $\alpha(t)$ satisfying,  for some
$0<q<1$,
\begin{equation}\label{eq:alpha^q}
\int_0^\infty \alpha(t)^q\,\dd t<\infty.
\end{equation}
\end{enumerate}
\end{cond}
\noindent
The kernel function $w$ is assumed to satisfy the following
conditions (an example of such a kernel is given in~(\ref{simkernel}) below, see also Wand~\cite{Wand98}),
which includes in particular the behavior of $\phi_{w}$ at the
boundary of its domain.
\begin{cond} \label{cond:w}
Let $w$ be a real symmetric function
with real valued symmetric characteristic function $\phi_w$ with support
[-1,1].
Assume further
\begin{enumerate}
\item
$\infint |w(u)|du < \infty$
,
$\infint w(u)du=1$
,
$\infint u^2|w(u)|du<\infty$
,
\item
$\phi_w(1-t)=At^\rho + o(t^\rho),\quad\mbox{as}\ t\downarrow 0 $
for some $\rho >0$, $A\in\rr$.
\end{enumerate}
\end{cond}

The first part of Condition~\ref{cond:sigma} is motivated by the
situation where $X=\sigma^2$ solves a SDE
like~(\ref{eq:sde}). It is easily verified that for such processes
it holds that $\ex |\sigma^2_t-\sigma^2_0| = O(t^{1/2})$, provided that $b \in
L_1(\mu)$ and $a \in L_{2}(\mu)$, where $\mu$ is the invariant
probability measure. Indeed we have  $\ex |\sigma^2_t-\sigma^2_0| \leq \ex
\int_0^t |b(\sigma^2_s)|\,ds + (\ex \int_0^t a^2(\sigma^2_s)\, ds)^{1/2} =
t||b||_{L_1(\mu)} + \sqrt{t} ||a||_{L_2(\mu)}$.

The main result we present for this estimator concerns its mean
squared error at a fixed point $x$. Although the motivation of the
estimator was based on the simplified model~(\ref{eq:sdesimple}),
the result below applies to the original model~(\ref{eq:sde}). For
its proof and additional technical details, see~Van Es et
al.~\cite{bert03}.

\begin{thm} \label{contasthm}
Assume that $\ex b_t^2$ is bounded. Let the process $\sigma$
satisfy Condition~\ref{cond:sigma}, and let the kernel function
$w$ satisfy Condition~\ref{cond:w}. Moreover, let the density $f$
of\/ $\log\sigma^2_t$ be twice continuously differentiable with a
bounded second derivative. Also assume that the density of
$\sigma^2_t$ is bounded in a neighbourhood of zero. Suppose that
$\Delta=n^{-\delta}$ for  given $0<\delta <1$ and choose
$h=\gamma\pi/\log n$, where $\gamma > 4/\delta$. Then the bias of
the estimator (\ref{eq:fnhc}) satisfies
\begin{equation}\label{contasthm:1}
\ex f_{nh}(x) - f(x) = \tfrac{1}{2}h^2f''(x)\int u^2w(u)du +
o(h^2),
\end{equation}
whereas, the variance of the estimator satisfies the order bounds
\begin{equation}\label{contasthm:2}
\var f_{nh}(x) =
O\Big(\frac{1}{n}\,h^{2\rho}e^{\pi/h}\Big)+O\Big(\frac{1}{{nh^{1+q}\Delta}}\Big).
\end{equation}
\end{thm}

\begin{rem}\label{remark:mse}
The choices $\Delta=n^{-\delta}$, with $0<\delta<1$ and
$h=\gamma\pi/\log n$, with $\gamma>4/\delta$ render a variance
that is of order $n^{-1+1/\gamma}(1/\log n)^{2\rho}$ for the first
term of (\ref{contasthm:2}) and $n^{-1+\delta}(\log n)^{1+q}$ for
the second term. Since by assumption $\gamma>4/\delta$ we have
$1/\gamma<\delta/4<\delta$ so the second term dominates the first
term. The order of the variance is thus $n^{-1+\delta}(\log
n)^{1+q}$. Of course, the order of the bias is logarithmic, hence
the bias dominates the variance and the mean squared error of
$f_{nh}(x)$ is of order $(\log n)^{-4}$.
\end{rem}
\noindent
\begin{rem}
It can then be shown that for the characteristic function $\phi_k$ one has the behavior
\[
|\phi_k(s)| = \sqrt{2}\,e^{-\frac{1}{2}\pi
|s|}(1+O(\tfrac{1}{|s|})),\,|s|\to\infty.
\]
This means that $k$ is supersmooth in the terminology of Fan~\cite{104} which explains the slow logarithmic rate at which the bias vanishes.
Sharper results on the variance can be obtained when $\sigma^2$ is
strongly mixing, see Van Es et al.~\cite{bert04} for further
details. The orders of the bias and of the MSE remain unchanged though.
\end{rem}
%\medskip\\
%If one wants to estimate a multivariate density, for
%instance the one of
%$(\log\sigma^2_{t_1},\ldots,\log\sigma^2_{t_p})$ for given
%$t_1<\cdots<t_p$, one uses a multivariate version of $f_{nh}$.
%This estimator has comparable results for bias and variance, see
%Van Es and Spreij~\cite{VanEsSpreij09}.

\subsection{An application to the Amsterdam AEX index}

\begin{figure}[h]
%$$ \epsfxsize=6cm\epsfysize=8cm\epsfbox{aexclose.pdf}
%\epsfxsize=6cm\epsfysize=8cm\epsfbox{logaexclose.pdf} $$
\begin{center}
\includegraphics[scale=0.7]{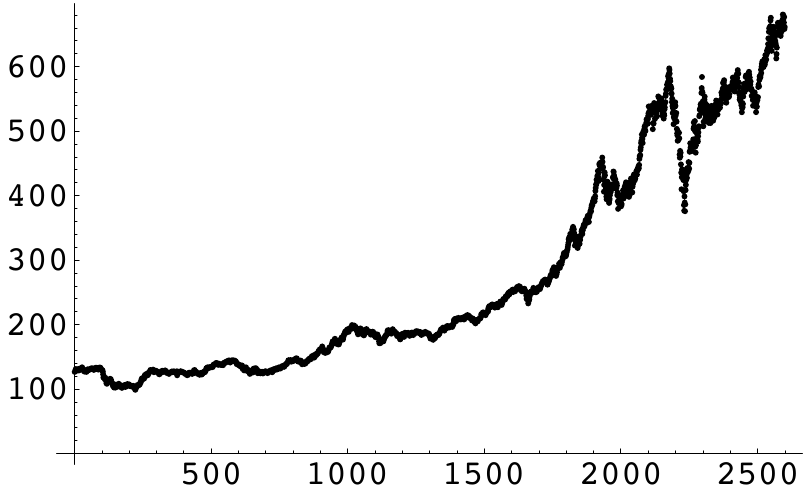}
\quad
\includegraphics[scale=0.7]{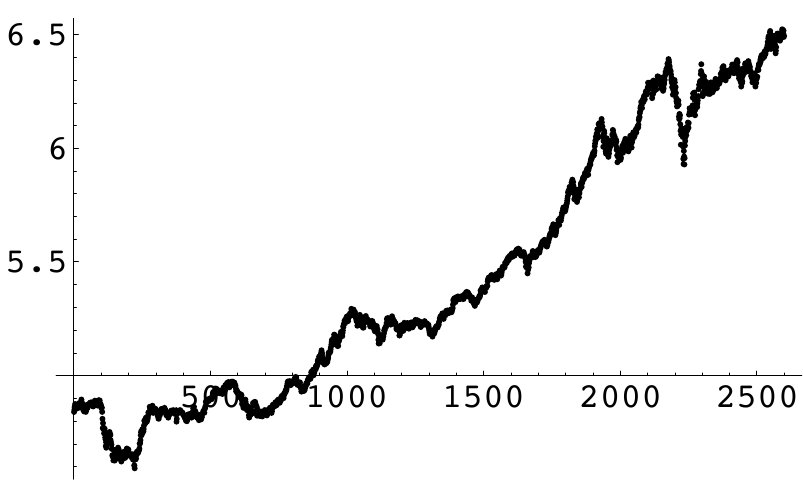}
\end{center}
\caption{AEX. Left: daily closing values. Right: log of the
daily closing values. \label{fig:10}}
\end{figure}

\begin{figure}[h]
%$$ \epsfxsize=6cm\epsfysize=8cm\epsfbox{meancorr-loglogreturns.pdf}
%\epsfxsize=6cm\epsfysize=8cm\epsfbox{meancorr-loglogreturnssq.pdf}
%$$
\begin{center}
\includegraphics[scale=0.7]{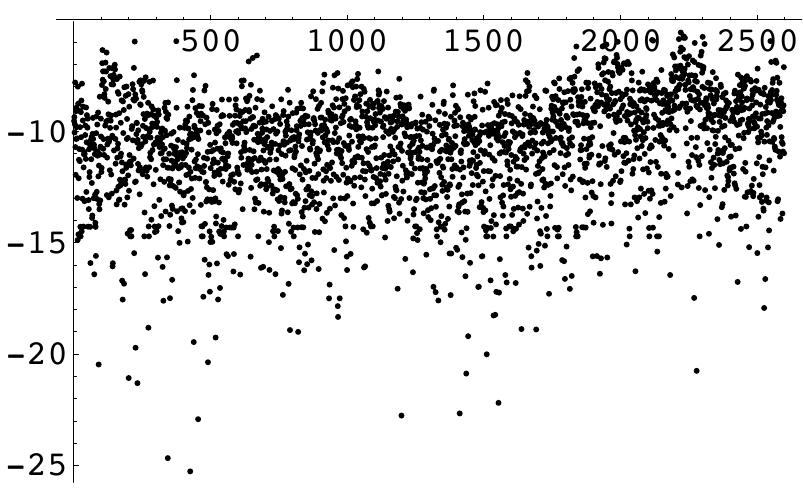}
\quad
\includegraphics[scale=0.7]{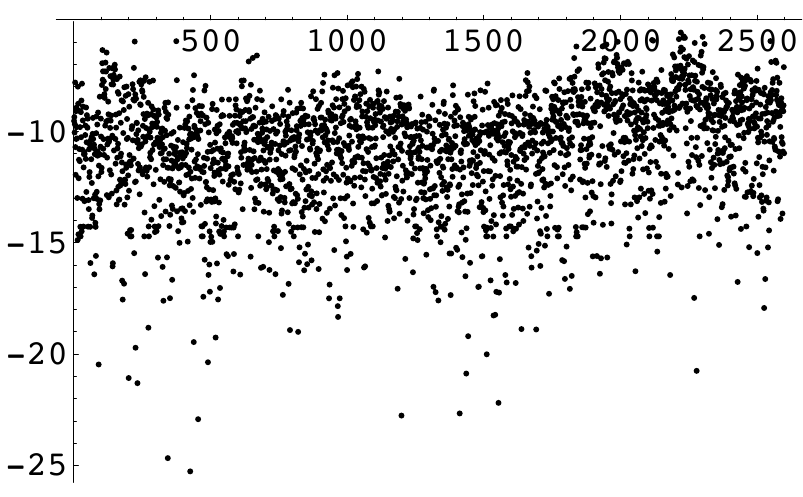}
\end{center}\caption{AEX. Left: the values of $X_t$, i.e. the centered daily
log returns. Right: $\log(X_t^2)$ . \label{fig:11}}
\end{figure}

\begin{figure}[h]
%$$ \epsfxsize=6cm\epsfysize=8cm\epsfbox{meancorr-aex-h07.pdf}
%\epsfxsize=6cm\epsfysize=8cm\epsfbox{meancorr-normalfit.pdf} $$
\begin{center}
\includegraphics[scale=0.7]{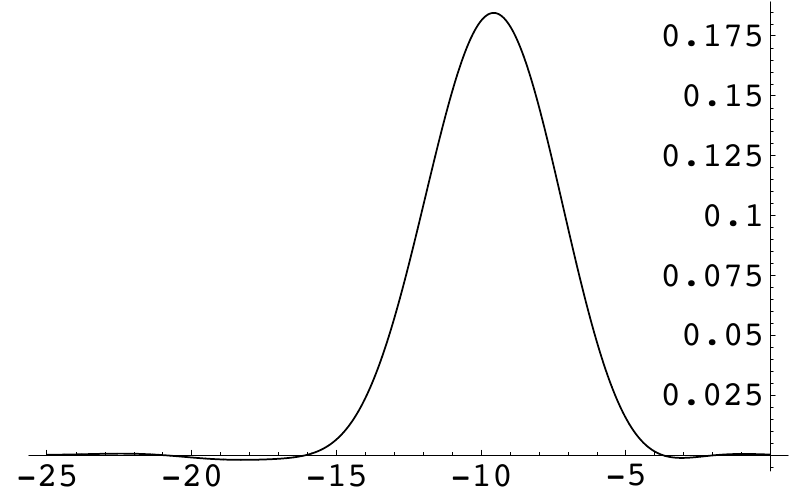}
\quad
\includegraphics[scale=0.7]{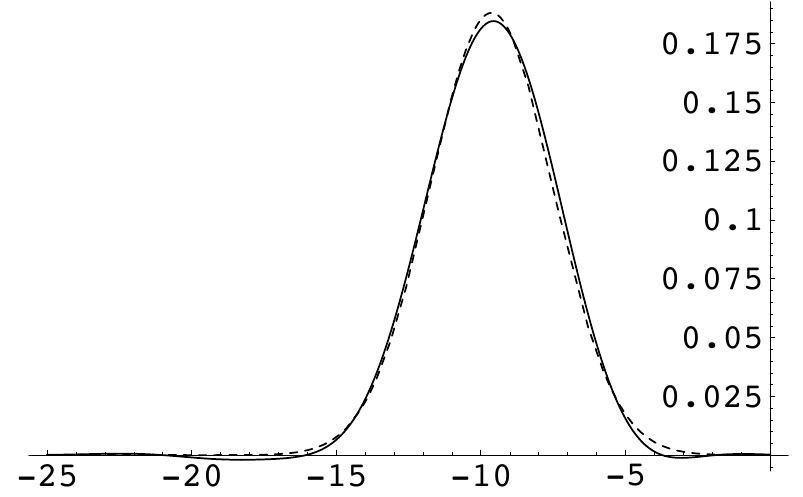}
\end{center}
\caption{AEX. Left: The estimate of the density of
$\log(\sigma_t^2)$ with $h=0.7$. Right: The normal fit to the
$\log(\sigma_t^2)$. The dashed line is the normal density and the
solid  line the kernel estimate.  \label{fig:13}}
\end{figure}

In this section we present an example using real data of the
Amsterdam AEX stock exchange. We have estimated the volatility
density from 2600 daily closing values of the Amsterdam stock
exchange index AEX from 12/03/1990 until 14/03/2000. These data
are represented in Figure~\ref{fig:10}. We have centered the daily
log returns, i.e we have subtracted the mean (which equaled
0.000636), see Figure~\ref{fig:11}. The deconvolution estimator is
given as the left hand picture in Figure~\ref{fig:13}. Observe
that the estimator strongly indicates that the underlying density
is unimodal. Based on computations of the mean and variance of the
estimate, with $h=0.7$, we have also fitted a normal density by
hand and compared it to the kernel deconvolution estimator. The
result is given as the right hand picture in Figure~\ref{fig:13}.
The resemblance is remarkable.

The kernel used to compute the estimates is a kernel from
Wand~\cite{Wand98}, with $\rho=3$ and $A=8$,
\begin{equation}\label{simkernel}
w(x)={\frac{48x(x^2-15)\cos x - 144(2x^2-5)\sin x}{\pi x^7}}.
\end{equation}
It has characteristic function
\begin{equation}\label{simchar}
\phi_w(t)=(1-t^2)^3,\quad |t|\leq 1.
\end{equation}
The bandwidths are chosen by hand. The estimates have been
computed by fast Fourier transforms using the Mathematica 4.2
package.

This is actually the same example as in our paper Van Es et
al.~\cite{bert04} on volatility density estimation for discrete
time models. The estimator (\ref{eq:fnhc})  presented here is, as
a function of the sampled data, exactly the same as the one for
the discrete time models. The difference lies in the choice of
underlying model. In the present paper the model is a discretely
sampled continuous time process, while in Van Es et
al.~\cite{bert04} it is a discrete time process. For the latter
type of models the discretization step in the beginning of this
section is not necessary since these models satisfy an exact
convolution structure.

\section{Wavelet deconvolution}\label{section:wavelet}

As an alternative to kernel methods, in this section we consider
estimators based on wavelets. Starting point is again the
simplified model~(\ref{eq:sdesimple}). Contrary to the previous
section, we are now interested in estimating the accumulated
squared volatility over an interval of length $\Delta$. We assume
having observations of $S$ at times $i\Delta$ to our disposal, but
now with $\Delta$ fixed (low frequency observations). Let, as
before,  $X^\Delta_i=\Delta^{-1/2}(S_{i\Delta} - S_{(i-1)\Delta})$
and let
$\bar{\sigma}_i^2=\Delta^{-1}\int_{(i-1)\Delta}^{i\Delta}\sigma^2_t\,\dd
t$. Denote by $\mathcal{F}_\sigma$ the $\sigma$-algebra generated
by the process $\sigma$. By the assumed independence of the
processes $\sigma$ and $W$, we have for the characteristic
function of $X^\Delta_i$ given $\mathcal{F}_\sigma$
\[
\EE [\exp(\ii sX^\Delta_i)| \mathcal{F}_\sigma]=\exp(-\half \bar{\sigma}_i^2s^2).
\]
Consider also the model $\tilde X^{\Delta}_i=\bar{\sigma_i}Z_i$, with $\bar{\sigma_i}$ and $Z_i$ independent for each $i$ and $Z_i$ a standard Gaussian random variable. Then
\[
\EE [\exp(\ii s\tilde X^{\Delta}_i)| \mathcal{F}_{\sigma_i}]=\exp(-\half \bar{\sigma}_i^2s^2).
\]
It follows that $X^\Delta_i$ and $\tilde X^{\Delta}_i$ are identically distributed. From this observation we conclude that the transformed increments
$\log\big(\Delta^{-1}(S_{i\Delta} - S_{(i-1)\Delta})^2\big)$ are then distributed as
$Y_i = \xi_i + \eps_i$, where
\[
\xi_i = \log \bar{\sigma}^2_i, \  \ \
\eps_i = \log Z_i^2,
\]
and $Z_i$ is an i.i.d.\ sequence of standard Gaussian random
variables, independent of $\sigma$. The sequence $\xi_i$ is
stationary and we assume that its marginal density $g$ exists,
i.e.\ $g$ is the density of
$\log \big(\Delta^{-1}\int_0^\Delta \sigma^2_u\,\dd u\big)$. The density of the $\eps_i$ is again denoted by $k$.
Of course, estimating $g$ is equivalent to estimating the
density of the aggregated squared volatility $\int_0^\Delta
\sigma^2_u\,du$.

In the present section the main focus is on the quality of the
estimator in terms of the mean integrated squared error, as
opposed to establishing results for the (pointwise) mean squared error as in
Section~\ref{section:kernel}. At the end of this section we
compare the results presented here to those of
Section~\ref{section:kernel}.

First we recall the construction of the wavelet estimator proposed
in Pensky and Vidakovic~\cite{103}. For the necessary background
on wavelet theory, see for instance Blatter~\cite{Bla98}, Jawerth
and Sweldens~\cite{102}, and the references therein.
For the construction of  deconvolution estimators we need to use
band-limited wavelets. As in Pensky and Vidakovic~\cite{103} we
use a Meyer-type wavelet (see also Walter~\cite{107}, Walter and
Zayed~\cite{108}). We consider an orthogonal scaling function and
wavelet $\varphi$ and $\psi$, respectively, associated with an
orthogonal multiresolution analysis of $L^2(\RR)$. We denote in this section the Fourier transform of a  function $f$ by
$\tilde f$, i.e.\
\[
\tilde f ( \omega )= \int_\RR e^{-\ii \omega x} f(x)\,\dd x,
\]
and suppose
that for a symmetric probability measure $\mu$ with support
contained in $\left[ - \pi/3, \pi/3 \right]$ it holds that
\[
  \tilde \varphi (\omega) = \Big(\mu(\omega-\pi, \omega+\pi] \Big)^{1/2}, \ \
 \tilde \psi (\omega) = e^{-i\omega/2} \Big(\mu(|\omega|/2-\pi, |\omega| -\pi]\Big)^{{1}/{2}}.
\]
Observe that the assumptions imply that $\varphi$ and $\psi$ are indeed band-limited. For the supports
of their Fourier transforms we have
  $\text{supp}\,\tilde\varphi \subset \left[ -4\pi/3,4\pi/3\right]$ and
$\text{supp}\,\tilde\psi \subset \left[-8\pi/3,-2\pi/3\right] \cup
\left[ 2\pi/3,8\pi/3\right]$. By choosing $\mu$ smooth enough we
ensure that $\tilde\varphi$ and $\tilde\psi$ are at least twice
continuously differentiable.

For any integer $m$, the unknown density  $g$ can now be written as
  \begin{equation}\label{213610}
  g(x)=\sum_{l\in {\mathbb Z}} a_{m,l}\varphi_{m,l}(x) +
                   \sum_{l\in {\mathbb Z}} \sum_{j=m}^{\infty} b_{j,l}\psi_{j,l}(x),
 \end{equation}
where
$\varphi_{m,l}(x) = 2^{m/2}\varphi (2^m x-l)$, $\psi_{j,l}(x) = 2^{j/2}\psi (2^jx-l)$
and the coefficients are given by
\[
a_{m,l}=\int_{\RR}\varphi_{m,l}(x)g(x)\, \dd x \, ,\quad   \quad
b_{j,l}=\int_\RR \psi_{j,l}(x)g(x)\,\dd x.
\]
The idea behind the linear wavelet estimator is simple. We first approximate $g$ by the
orthogonal projection given by the first term on the right-hand side of (\ref{213610}).
For $m$ large enough the second term will be small, and can be controlled by using the
approximation properties of the specific family of wavelets that is being used.
The projection of $g$ is estimated by replacing the coefficients $a_{m,l}$ by consistent
estimators and truncating the sum.
Using the fact that the density $p$ of an observation $Y_i$ is the convolution of $g$ and $k$ it is
easily verified that
\[
             a_{m,l}=\int_\RR2^{m/2}U_m(2^mx-l)p(x)\,\dd x =  2^{m/2}\EE U_m(2^mY_i-l),
\]
where $U_m$ is the function with Fourier transform
\begin{equation}\label{eq: um}
\tilde U_m (\omega) = \frac{\tilde \varphi (\omega) }{\tilde k(-2^m \omega)}.
\end{equation}
We estimate the coefficient
 $a_{m,l}$ by its empirical counterpart
\[
\hat a_{m,l,n}=\frac{1}{n} \sum_{i=1}^{n} 2^{{m}/{2}} U_m(2^mY_i-l).
 \]
Under the mixing assumptions that we will impose on the sequence
$Y$, it will be stationary and ergodic. Hence, by the ergodic
theorem, $\hat a_{m,l,n}$ is a consistent estimator for $a_{m,l}$.
The  wavelet estimator is now defined by
 \begin{equation}\label{213619}
 \hat g_{n}(x)= \sum_{\left|l\right| \le L_n} \hat a_{m_n,l,n} \varphi_{m_n,l}(x),
 \end{equation}
where the detail level $m_n$ and the truncation point $L_n$ will
be chosen appropriately later.

The main results in the present section are upper bounds for the
mean integrated square error of the  wavelet estimator $\hat
g_{n}$, which is  defined as usual by
\[
\text{MISE}\left( \hat g_{n} \right)=\EE \int_{\RR} \left( \hat
g_{n} (x) - g(x) \right)^2\, \dd x.
\]
We will specify how to choose the detail level $m_n$ and the
truncation point $L_n$ in (\ref{213619}) optimally in different
cases, depending on the smoothness of $g$ and $k$. The smoothness
properties of $g$ are described in terms of $g$ belonging to
certain Sobolev balls and by imposing a weak condition on its
decay rate. The
Sobolev space $H^\alpha$ is defined for $\alpha > 0$ by
\begin{equation}\label{eq:sobolev}
H^{\alpha}= \Big\{g : \;\, \|g\|_{\alpha}=\Big( \int_\RR |\tilde
  g(\omega)|^2(\omega^2+1)^{\alpha}\,\dd \omega \Big)^{1/2} \! <\infty  \Big\}.
\end{equation}
Roughly speaking, $g \in H^\alpha$ means that the first $\alpha$
derivatives of $g$ belong to $L^2(\RR)$. The Sobolev ball of
radius $A$ is defined by
\[
\mathscr{S}_{\alpha}(A)=\left\{ g \in
H^{\alpha}: \, \|g\|_{\alpha} \le A\right\}.
\]
The additional assumption on the decay rate is reflected by $g$ belonging to
\[
\mathscr{S}^*_{\alpha}(A, A')=\mathscr{S}_{\alpha}(A) \cap
\Big\{g: \sup_x |xg(x)| \le A'\Big\}.
\]
We now have the following result, see Van Zanten and
Zareba~\cite{VanzantenZareba08}, for the wavelet density estimator
$\hat g_n$ of $g$ defined by (\ref{213619}).

\begin{thm}\label{thm: sv}
Suppose that the volatility process $\sigma^2$ is strongly mixing with mixing coefficients
satisfying
\begin{equation}\label{eq: karel}
\sum_{k \ge 0} \alpha_{k\Delta}^p < \infty
\end{equation}
for some $p \in (0,1)$. Then
 with the choices
\[
2^{m_n} = \frac{\log n}{1+(4\pi^2/3)}, \ \ \
L_n = \Big({\log n}\Big)^{r}, \ \ \ r \ge 1+2\alpha
\]
the mean square error of the wavelet estimator satisfies
\[
\sup_{g \in \mathscr{S}^*_{\alpha}(A, A')} \mathrm{MISE} \left( \hat g_{n} \right)
= O\Big({(\log n)^{-2\alpha}}\Big)
\]
for $\alpha, A, A' > 0$. If (\ref{eq: karel}) is satisfied for {\em all} $p \in (0,1)$,
the same bound is true if the choice for $L_n$ is replaced by $L_n = n$.
\end{thm}
Let us point out the relation with the results of Section~\ref{section:kernel} and with those in Van Es et
al.~\cite{bert04}, see also Section~\ref{section:dtm}. In that
paper kernel-type deconvolution estimators for discrete time
stochastic volatility models were considered. When applied to the
present model the results say that under the same mixing condition
and assuming that $g$ has two bounded and continuous derivatives,
the (pointwise) mean squared error of the kernel estimator is of
order $(\log n)^{-4}$. The analogue of $g$ having two bounded
derivatives in our setting is that $g \in \mathscr{S}^*_{2}(A,
A')$ for some $A,A' > 0$. Indeed, the theorem yields the same
bound $(\log n)^{-4}$ for the MISE in this case. The same bound is
valid for the MSE when estimating the marginal density for
continuous time models,  see Theorem~\ref{contasthm} and its
consequences in Remark~\ref{remark:mse}. Theorem \ref{thm: sv} is
more general, because the smoothness level is not fixed at
$\alpha=2$, but allows for different smoothness levels of order
$\alpha\neq 2$ as well. Moreover, the wavelet estimator is adaptive in
the sense that it does not depend on the unknown smoothness level,
if the condition on the mixing coefficients holds for all $p \in
(0,1)$.

\section{Penalized projection estimators}\label{section:ppe}

The results of the preceding sections assume that the true
(integrated) volatility density has a finite degree of regularity,
either in H\"older or in Sobolev sense. Under this assumption the
nonparametric estimators  have logarithmic convergence rates, cf.\
Remark \ref{remark:mse} and Theorem \ref{thm: sv}. Although
admittedly slow, the minimax results of Fan~\cite{104} show that
these rates are in fact optimal in this setting. In the paper
Pensky and Vidakovic~\cite{103} it was shown however that if in a
deconvolution setting the density of the unobserved variables has
the same degree of smoothness as the noise density, the rates can
be significantly improved, cf.\ also the lower bounds obtained in
Butucea~\cite{but1} and Butucea and Tsybakov~\cite{but2}. This
observation forms the starting point of the paper Comte and
Genon-Catalot~\cite{com}, in which a nonparametric volatility
density estimator is developed that achieves better rates than
logarithmic if the true density is super smooth.

In
%Comte and Genon-Catalot~\cite{com}
the latter paper it is assumed that there are
observations $S_\Delta, S_{2\Delta}, \ldots, S_{n\Delta}$ of a
process $S$ satisfying the simple equation (\ref{eq:sdesimple}),
with $V = \sigma^2$ a $(0,\infty)$-valued process independent of
the Brownian motion $W$. It is assumed that we deal with high frequency observations, $\Delta\to 0$ and $n\Delta\to\infty$. We impose the following condition on $V$.
\begin{cond}\label{cond:v}
The process $V$ is a time-homogenous, continuous Markov process, strictly
stationary and ergodic.
It is either $\beta$-mixing with coefficient $\beta(t)$ satisfying
\[
\int_0^\infty \beta(t)\,\dd t < \infty,
\]
or is $\rho$-mixing.
Moreover, it
satisfies the Lipschitz condition
\[ \EE\Big( \log\Big(\frac1\Delta
\int_0^\Delta V_t\,\dd t\Big) - \log V_0\Big)^2 \le C\Delta,
\]
for some $C > 0$.
\end{cond}
\noindent In addition to this a technical assumption is necessary on the
density $f$  of $\log V_0$ we are interested in and on  the
density $g_\Delta$ of $\log\Big(\frac1\Delta \int_0^\Delta
V_t\,\dd t\Big)$, which is assumed to exist. Contrary to the notation of the previous section, we write $g_\Delta$ instead of $g$, since now $\Delta$ is not fixed.

\begin{cond}\label{cond:t}
The invariant density $f$ is bounded and has a second moment and
$g_\Delta \in L^2(\RR)$.
\end{cond}

\noindent As a first step in the construction of the final estimator a
preliminary estimator $\hat f_L$ is constructed for $L \in \NN$
fixed. Note that Condition~\ref{cond:t} implies that $f \in
L^2(\RR)$, hence we can consider its orthogonal projection $f_L$
on the subspace $S_L$ of $L^2(\RR)$, defined as the space of
functions whose Fourier transform is supported on the compact
interval $[-\pi L, \pi L]$. An orthonormal basis for the  latter
space is formed by the Shannon basis functions $\psi_{L,j}(x) =
\sqrt{L}\psi(Lx-j)$, $j  \in \ZZ$, with $\psi(x) = \sin(\pi
x)/(\pi x)$ the sinc kernel. For integers $K_n \to \infty$ to be
specified below, the space $S_L$ is approximated by the
finite-dimensional spaces $S^n_L = {\rm span}\{\psi_{L, j}: |j|
\le K_n\} $. The function $f_L$ is estimated by $\hat f_L =
\mathop{\rm argmin}_{h \in S^n_L}\gamma_n(h)$, where the contrast
function $\gamma_n$ is defined for $h\in L^2(\rr)\cap L^1(\rr)$ by
\[
\gamma_n(h)  = \|h\|^2_2 - \frac2n\sum_{i=1}^n u_h(\log
(X^\Delta_i)^2), \qquad u_h(x) = \frac1{2\pi}\int_{-\infty}^\infty
e^{\ii xs}\frac{\tilde h(-s)}{\phi_k(s)}\,\dd s.
\]
Here, as before, $\phi_k$ is the characteristic function of
$\log\eps^2$, with $\eps$ standard normal and $\tilde{h}$ is the Fourier transform of $h$. It is easily seen that
\[
\hat f_L = \sum_{|j| \le K_n} \hat a_{L,j}\psi_{L,j}, \qquad \hat
a_{L,j} = \frac1n\sum_{j=1}^n u_{\psi_{L,j}}(\log (X^\Delta_i)^2).
\]
Straightforward computations show that, with $\qv{\cdot,\cdot}$
the $L^2(\RR)$ inner product, $\EE u_h(\log (X^\Delta_i)^2) =
\qv{h,g_\Delta}$, and hence $\EE \gamma_n(h) = \|h-g_\Delta\|^2_2
- \|g_\Delta\|^2_2$. So in fact, $\hat f_L$ is an estimator of the
element of $S^n_L$ which is closest to $g_\Delta$. Since $S^n_L$
approximates $S_L$ for large $n$ and $g_\Delta$ is close to $f$
for small $\Delta$, the latter element should  be close to $f_L$.

Under Conditions~\ref{cond:v} and~\ref{cond:t},  a bound for the
mean integrated square error, or quadratic risk MISE$(\hat f_L) =
\EE\|\hat f_L - f\|^2_2$ can be derived,
% in the paper Comte and
%Genon-Catalot~\cite{com},
depending on the approximation error
$\|f-f_L\|_2$, the bandwidth $L$ and the truncation point $K_n$,
see Comte and Genon-Catalot~\cite{com}, Theorem 1. The result
implies that if $f$ belongs to the Sobolev space $H^\alpha$ as defined in~(\ref{eq:sobolev}), then the choices $K_n = n$ and $L
= L_n \sim \log n$ yield a MISE of the order $(\log n
)^{-2\alpha}$, provided that $\Delta = \Delta_n = n^{-\delta}$ for
some $\delta \in (0,1)$. Not surprisingly, this is completely
analogous to the result obtained in Theorem \ref{thm: sv} for the
wavelet-based estimator in the fixed $\Delta$ setting. In
particular the procedure is adaptive, in that the estimator does
not depend on the unknown regularity parameter $\alpha$.

To obtain faster than logarithmic rates and adaptation in the case
that $f$ is supersmooth, a data-driven choice of the bandwidth $L$
is proposed. Define
\[
\hat L = \mathop{\rm argmin}_{L \in \{1, \ldots, \log n\}}
\Big(\gamma_n(\hat f_L) + {\rm pen}_n(L)\Big),
\]
where the penalty term is given by
\[
{\rm pen}_n(L) = \kappa \frac{(1+L)\Phi_k(L)}{n},
\]
for $\kappa > 0$ a calibration constant and
\[
\Phi_k(L) = \int_{-\pi L}^{\pi L}\frac{1}{|\phi_k(s)|^2}\,ds.
\]
For the quadratic risk of the estimator $\hat f_{\hat L}$, the
following result holds (Comte and Genon-Catalot~\cite{com}).

\begin{thm}
Under Conditions~\ref{cond:t} and~\ref{cond:v} we have
\begin{align*}
{\rm MISE}(\hat f_{\hat L}) & \le C_1 \inf_{L \in \{1, \ldots, \log n\}} \Big(\|f-f_L\|^2_2 + \frac{(1+L)\Phi_k(L)}{n}\Big) \\
& \quad + C_2\frac{\log^2 n}{K_n} + C_3\frac{\log n}{n\Delta} +
C_4\Delta \log^3 n,
\end{align*}
for constants $C_1, C_2, C_3, C_4 > 0$.
\end{thm}
It can be seen that  this bound is worse than  the corresponding
bound for the estimator $\hat f_L$  by a factor of the order $L$.
This is at worst a logarithmic factor which, as usual in this kind
of setting, has to be paid for achieving adaptation. The examples
in Section 6 of Comte and Genon-Catalot~\cite{com} show that
indeed, the estimator $\hat f_{\hat L}$ can achieve algebraic
convergence rates in case the true density $f$ is supersmooth.

\section{Estimation for discrete time models}\label{section:edtm}

Although the main focus of the present paper is on estimation procedures for continuous time models, in the present section we also highlight some analogous results for discrete time models. These deal with both density and regression function estimation.

\subsection{Discrete time models}\label{section:dtm}

The discrete analogue of~(\ref{eq:sdesimple}) is
\begin{equation}\label{eq:s}
X_t = \sigma_t Z_t,\,t=1,2,\ldots .
\end{equation}
Here we denote by  $X$  the detrended or demeaned log-return
process. Stochastic volatility models are often described in this form.
The sequence $Z$ is typically an i.i.d.\ noise (e.g.\ Gaussian)
and at each time $t$ the random variables $\sigma_t$ and $Z_t$ are
independent. See the survey papers by Ghysels et al.~\cite{GHR96}
or Shephard~\cite{Shephard96}. Also in this section we assume that
the process $\sigma$ is strictly stationary and  that the marginal
distribution of $\sigma$  has a density with respect to the
Lebesgue measure on $(0,\infty)$. We present some results for a
nonparametric estimator of the   density of $\log \sigma^2_t$, as
well as results for a nonparametric estimator of a nonlinear
regression function, in case $\sigma^2$ is given by a
nonlinear autoregression. The standing assumption in all what
follows is that for each $t$ the random variables $\sigma_t$ and
$Z_t$ are independent, the noise sequence is standard Gaussian and
$\sigma$ is a strictly stationary, positive process satisfying a
certain mixing condition.

In principle one can distinguish two classes of models. The way in
which the bivariate process $(\sigma,Z)$, in particular its
dependence structure, is further modelled offers different
possibilities. In the first class of models one assumes that the
process $\sigma$ is predictable with respect to the filtration
$\FF_t$ generated by the process $Z$, and obtains that $\sigma_t$
is independent of $Z_t$ for each fixed time $t$. We furthermore
have that (assuming that the unconditional variances are finite)
$\sigma^2_t$ is equal to the conditional variance of $X_t$ given
$\mathcal{F}_{t-1}$. This class of models has become quite popular
in the econometrics literature.  It is well known that  this class
also contains the (parametric) family of GARCH-models, introduced
by Bollerslev~\cite{Bollerslev86}.

In the second  class of models one assumes that the whole process
$\sigma$ is independent of the noise process $Z$, and one commonly
refers to the resulting model as a stochastic volatility model. In
this case, the natural underlying filtration
$\FF=\{\mathcal{F}_t\}_{t\geq 0}$ is generated by the two
processes $Z$ and $\sigma$ in the following way. For each $t$ the
$\sigma$-algebra $\mathcal{F}_t$ is generated by $Z_s$, $s\leq t$
and $\sigma_s$, $s\leq t+1$. This choice of the filtration
enforces $\sigma$ to be predictable. As in the first model the
process $X$ becomes a martingale difference sequence and we have
again (assuming that the unconditional variances are finite) that
$\sigma^2_t$ is the conditional variance of $X_t$ given
$\mathcal{F}_{t-1}$.
 An example of such a model is given in De Vries~\cite{DeVries98}, where
$\sigma$ is generated as an AR(1) process with $\alpha$-stable
noise ($\alpha\in(0,1)$).

As in the previous sections we refrain from parametric modelling
and review some completely nonparametric approaches. We will
mainly focus on results for the second class, as it is the
discrete time analogue of the stochastic volatility models of the
previous sections. At the heart of all what follows is again the
convolution structure that is obtained from~(\ref{eq:s}) by
squaring and taking logarithms,
\[
\log X_t^2 = \log \sigma_t^2 + \log Z_t^2.
\]

\subsection{Density estimation}

The main result of this section gives a bias expansion and a
variance bound of a kernel density type estimator of the  density
$f$ of  $\log\sigma^2_t$, which chosen to be, analogously
to~(\ref{eq:fnhc}),
\begin{equation}\label{eq:fnhd}
f_{nh}(x)=\frac{1}{nh}\sum_{j=1}^n v_h\left(\frac{x-\log
(X_j)^2}{h}\right),
\end{equation}
where $v_h$ is  the kernel function of~(\ref{fourkernel}).

The next theorem is derived from Van Es et al.~\cite{bert04},
where a multivariate density estimator is considered. It
establishes the expansion of the bias and an order bound on the
variance of our estimator under a strong mixing condition. Under
broad conditions this mixing condition is satisfied if the process
$\sigma$  Markov, since then convergence of the mixing
coefficients to zero takes place at an {\em exponential rate}, see
Theorems 4.2 and Theorem 4.3 of Bradley~\cite{Bradley86}
 for precise statements. Similar behaviour occurs for ARMA processes
with absolutely
 continuous distributions of the noise terms~(Bradley~\cite{Bradley86}, Example
6.1).

\begin{thm}\label{discrasthmp}
Assume that the process $\sigma$ is strongly mixing with
coefficient $\alpha_k$ satisfying $$ \sum_{j=1}^\infty
\alpha_j^{\beta}<\infty, $$ for some $\beta\in (0,1)$. Let the
kernel function $w$ satisfy Condition~\ref{cond:w} and let the density $f$ of
  $\log\sigma^2_t$ be
bounded and twice continuously differentiable with bounded second
order partial derivatives. Assume furthermore that $\sigma$ and
$Z$ are  independent processes. Then we have for the estimator of
$f$ defined as in~(\ref{eq:fnhd}) and $h\to 0$
\begin{equation}\label{discrasthm:1p}
\ex f_{nh}(x)= f(x)+\tfrac{1}{2}h^2f''(x)\int u^2 \, w(u)\,\dd
u+o(h^2)
\end{equation}
and
\begin{equation}\label{discrasthm:3p}
\var f_{nh}(x) = O\big(\tfrac{1}{n}\,h^{2\rho}\,e^{\pi/h}\big).
\end{equation}
\end{thm}

\begin{rem}
Comparing the above results to the ones in
Theorem~\ref{contasthm}, we observe that in the continuous time
case, the variance has an additional
$O\Big(\frac{1}{{nh^{1+q}\Delta}}\Big)$ term.
\end{rem}

\subsection{Regression function estimation}

In this section we  assume the basic model~(\ref{eq:s}), but in
addition we assume that the process $\sigma$ satisfies a nonlinear
autoregression and we consider nonparametric estimation of
the regression function as proposed in Franke et al.~\cite{FHK03}.
In that paper a discrete time model was proposed as a
discretization of the continuous time model given
by~(\ref{eq:sde}). In fact, Franke et al.\ include a mean
parameter $\mu$, but since they assume it to be known, without
loss of generality we can still assume~(\ref{eq:s}). Assume that
the volatility process is strictly positive and consider $\log
\sigma_{t}^2$. It is assumed that its evolution is governed by
\begin{equation}\label{eq:ar}
\log\sigma_{t+1}^2=m(\log\sigma_{t}^2)+\eta_t,
\end{equation}
where the $\eta_t$ are i.i.d.\ Gaussian random variables with zero
mean. The regression function $m$ is assumed to satisfy the
stability condition
\begin{equation}\label{eq:mstable}
\limsup_{|x|\to\infty}|\frac{m(x)}{x}|<1.
\end{equation}
Under this condition the process $\sigma$ is exponentially ergodic
and strongly mixing, see Doukhan~\cite{Doukhan94} and these
properties carry over to the process $X$ as well. Moreover, the
process $\log \sigma^2_t$ admits an invariant density $f$.

Denoting $Y_t=\log X_t^2$, we have
\[
Y_t=\log\sigma_{t}^2+\log Z_t^2.
\]
It is common to assume that the processes $Z$ and $\eta$ are
independent, the second class of models described in
Section~\ref{section:dtm}, but dependence between $\eta_t$ and
$Z_t$ for fixed $t$ can be allowed for  (first model class)
without changing in what follows, see Franke et al.~\cite{FHK03}.

The purpose of the present section is to estimate the function $m$
in~(\ref{eq:ar}). To that end we use the estimator $f_{nh}$ as
defined in (\ref{eq:fnhd}). Since this estimator resembles an
ordinary kernel density estimator, the important difference being
that the kernel function $v_h$ now depends on the bandwidth $h$,
the idea is to mimic the classical Nadaraya-Watson regression
estimator similarly, in order to obtain an estimator of $m(x)$.
Doing so, one obtains the estimator
\begin{equation}\label{eq:estim}
m_{nh}(x)= \frac{\frac{1}{nh}\sum_{j=1}^n v_h\big(\frac{x-Y_j}{h}\big)Y_{j+1}}{f_{nh}(x)}.
\end{equation}
It follows that
\[
m_{nh}(x)-m(x)=\frac{p_{nh}(x)}{f_{nh}(x)},
\]
where
\[
p_{nh}(x)=\frac{1}{nh}\sum_{j=1}^n v_h\big(\frac{x-Y_j}{h}\big)(Y_{j+1}-m(x)).
\]
In Franke et al.~\cite{FHK03} bias expansions for $p_{nh}(x)$ and
$f_{nh}$ are given that fully correspond to those in
Theorem~\ref{discrasthmp}. They are again of order $h^2$, under
similar assumptions. It is also shown that the variances of
$p_{nh}$ and $f_{nh}$ tend to zero. The main result concerning the
asymptotic behavior then follows from combining the asymptotics
for $p_{nh}$ and $f_{nh}$.
\begin{thm}
Assume that $m$ satisfies the stability
condition~(\ref{eq:mstable}), that $m$ and $f$ are twice
differentiable and the first of Condition~\ref{cond:w} on the
kernel $w$. The estimator $m_{nh}(x)$ satisfies $(\log
n)^2(m_{nh}(x)-m(x))=O_p(1)$ if $h=\gamma/\log n$ with
$\gamma>\pi$.
\end{thm}
Following the proofs in Franke et al.~\cite{FHK03}, one can
conclude that e.g. the variance of $p_{nh}$ is of order
$O(\frac{\exp(\pi/h)}{nh^4})$, which tends to zero for
$h=\gamma/\log n$, with $\gamma>\pi$. For the variance of $f_{nh}$
a similar bound holds. Comparing these order bounds to the ones in
Theorem~\ref{discrasthmp}, we see that the latter ones are
sharper. This is partly due to the fact that Franke et
al.~\cite{FHK03}, don't impose conditions on the boundary behavior
of the function $\phi_w$ (the second of Condition~\ref{cond:w}),
whereas their other assumptions are the same as in Theorem~\ref{discrasthmp}.

\section{Concluding remarks}\label{section:remarks}

In recent years, many different parametric stochastic volatility
models have been proposed in the literature. To investigate which
of these models are best supported by observed asset price data,
nonparametric methods can be useful. In this paper we reviewed a
number of such methods that have recently been proposed. The
overview shows that ideas from deconvolution theory can  be
instrumental in dealing with this statistical problem and that
both for high and for low frequency data, methods are now
available for nonparametric estimation of the (integrated)
volatility density at optimal convergence rates.

On a critical note, the methods available so far all assume that
the volatility process is independent of the Brownian motion
driving  the asset price dynamics. This is a limitation, since in
several interesting models non-zero correlations are assumed
between the Brownian motions driving the volatility dynamics and
the asset price dynamics.
%This point clearly deserves further
%attention in the future. It is closely connected to deconvolution
%problems in which the observed signal and the noise are allowed to
%be dependent. [DOES THIS MAKE ANY SENSE AT ALL?]

\end{document}